# Vapor-liquid equilibrium (VLE) properties versus critical exponent theory – on new approximate mathematical approach to determine the critical exponent value of the vapor-liquid coexistence curve

Beata Staśkiewicz [1], Robert Stańczy[2,*]

[1]Wrocław University of Technology, Wybrzeże Wyspiańskiego 27, 50-370 Wrocław, Poland

[2]University of Wrocław, pl. Grunwaldzki 2/4, 50-384 Wrocław, Poland

A novel mathematical rigorous method to obtain the vapor-liquid equilibrium curves near the critical point has been proposed covering the cases of well-known and commonly used equations of state for real gases - Van der Waals and Dieterici, which are prototypes to determine the critical exponent value of the coexistence curves system. We explain that the critical exponent value of the vapor-liquid equilibrium curves can be regarded as the analogy to the results based on classical assumptions – *mean field theory*. The novelty of our method has been exposed in comparison to the standard thermodynamic limit discussed thoroughly and extensively in the literature.

## 1. Introduction

It is impossible not to notice that in the Nature changes of phase (generally called *phase transitions*) are ubiquitous phenomena not just in natural processes but also in real world applications. The phase transition, which is the most important and fundamental, is the one between the liquid and vapor where there is phase of a pure fluid. Towards the end of the last century when the extensive studies of Andrews [1] have been used by Van der Waals (VdW) [2] to explain the influence of the temperature and pressure on the fluid's density. Thus the first theoretical description of the vapor-liquid phase transition with Van der Waals equation of state (VdW EOS) was available. Known is the fact that according to the Gibbs phase rule thermodynamic equilibrium of a pure component can be expressed as a three different phases. Each of them is determined by only two independent state variables (as for example $P$ – pressure, $\rho$ – mass density, $T$ – temperature). If we imagine the projection of the state surface on the plane ($P$, $\rho$) *i.e.* pressure and mass density, we see that the states as solid, liquid and gas (vapor) are represented separated by coexistence curve. In the case when these state variables of a pure component are located on the coexistence curve, then this component is observed as a two coexisting phases. The top of the vapor-liquid coexistence curve is named the critical point. We must note in this place that in this work we do not consider a fluids whose temperature is higher than critical temperature. These fluids are named supercritical fluids [3] because they exhibit atypical properties, at times related to those of gases and at times of those of fluids.

As it was mentioned above in the field of the phase transitions one of the most interesting and of key importance is the critical point. This is the point (reference point) at which the distinction between the liquid and gas phase just disappears and from which all fluids properties can be derived. The critical point is characterized by fixed temperature, pressure and density and also is an example of a continuous phase transition in the case of the vapor – liquid system [4]. It is so, because the vapor – liquid critical point is one of the many critical points associated with transitions where density (as a one of its specific thermodynamic property) changes continuously in the course of transition. According to the hypothesis that all properties of critical points are due to scale invariance of critical phases, thermodynamic properties of the system exhibit either divergencies or extinctions as stated by quite simple mathematical function. As a rule these functions are called power laws with well-defined critical exponents. It is worth noting that very often the problem of the phase transition is considered from the point of view of statistical mechanics, seeking an explanation in terms of macroscopic properties. Based on this approach VdW EOS has been obtained by method which is more general (in contrast to the ones based on calculation of virial equation of state up to second order and resummation of the hardsphere contribution to the pressure and direct use of perturbation theory) in the sense that it contains the idea of the *mean field* which pervades the field of statistical mechanics.

---

* Author e–mail address: stanczr@math.uni.wroc.pl

With the introduction in the Landau theory the order parameter [5], the behavior of any thermodynamic property which is derived from Helmholz free energy, is universal. Moreover, the Landau expansion as well as VdW EOS provide description of the critical point, called "*classical*" or equivalently *mean field theory*. Based on this, the phase transition of any system can be expressed by an analytic formula (for symmetry reasons only the even powers of $\tilde{\rho}$ are considered in this formula),

$$F(T,\tilde{\rho}) = F_0(T,0) + \frac{a}{2}(T - T_c)\tilde{\rho}^2 + \frac{b}{4}\tilde{\rho}^4 + \cdots \tag{1}$$

where $F$ – Helmholz free energy, $T$ – temperature, $\tilde{\rho}$ – order parameter (for vapor – liquid equilibrium system corresponds to the dimensionless density difference, denoted as $\tilde{\rho} = (\rho - \rho_c)/\rho_c$, and $\rho_c$ is the critical density in this case).

Minimizing $F$ one can obtain the equilibrium state of the vapor-liquid system

$$\frac{\partial F}{\partial \tilde{\rho}} = 0; \qquad \tilde{\rho}[a(T - T_c) + b\tilde{\rho}^2 + \cdots] = 0 \tag{2}$$

which has approximated two solutions for the case when $T < T_c$, given by

$$\tilde{\rho}^{\pm} = \pm \left(\frac{a}{b}\right)^{1/2} (T - T_c)^{1/2} \tag{3}$$

The solution of Eq. (3) in general case describes the coexistence curve of vapor-liquid equilibrium (VLE) system.

In most cases used asymptotical power laws and correlated with them order parameters are in real the first terms so called Wegner expansion near the critical point [6]. This expansion accounts for the correction to scaling outside the range of the power law behavior and for fluids is in the following form,

$$\Delta K = K_0 |t|^{\theta_0}[1 + K_1 |t|^{\Delta_1} + K_2 |t|^{2\Delta_1} + \cdots] \tag{3a}$$

where $\Delta K$ – quantity diverging at critical temperature $T_c$, $K_0$ – leading amplitude term, $K_i$ – corresponds to correction amplitudes or coefficients, $t = (T/T_c - 1)$ – deviation parameter ('distance' from the critical point) in the case of the temperature (see also Sec.3), $\theta_0$ – critical exponent, $\Delta_1$ – correction exponent. It must be underlined that correction amplitudes and correction exponent are also known as confluent singularities.

In light of our current understanding vapor-liquid critical phenomena in the framework of renormalization groups (RG) has been analyzed in [7]. It is the theory which gives just an imperfect and partial description the behavior of various fluid properties in asymptotic neighborhood of the vapor–liquid critical point, *i.e.* limiting exponents and ratios of amplitudes and fails to the amplitudes themselves (for more details see also Tabs. 1 and 2) [7-8]. Usually the values of the critical amplitudes are strictly dependent on the system analyzed and the substances tested (Tab. 3) [9-10]. This paper proposes approximate mathematical as well as quantitative and qualitative way to determine not only the value of the critical exponent for the vapor-liquid system (see method described below) but also its value of the critical amplitude, in the general case [9-10]). From theoretical point of view the 'universal' value of the critical amplitudes may be a consequence of the use of the law of corresponding states, in which if the reduced intermolecular potentials are identical for a set of substances, all of the reduced properties including the critical parameters and critical amplitudes of those substances, will be the same, for more details see [10] and References therein.

Up to now one can find in the literature many examples or many theories, which allow determine the value of critical exponent by different way, one can distinguish among them theoretical methods, generally based on theory of renormalization groups (RG) [7, 11-12]; refer to advanced methods of functional analysis or quantum mechanics [13-14] as well as experimental and molecular simulation efforts [15-22], both for single and binary or ternary systems (as for example work of Van Konynenburg, which detailed describe critical behavior of binary systems for VdW EOS [23]). Analytical derivation value of the critical exponent for the VdW EOS one can find also in the work of Reichl [24].

This paper throw more light on a new approximate mathematical approach to the problem of determining the value of critical exponent for the VLE system and hence for a new description of the coexistence curve. We pay a great attention to the method derived by Okrasiński et al. [25] and modified by Staśkiewicz et al.[26] to enable the analysis of the vapor-liquid equilibrium system in the vicinity of the critical point. In this paper as a first approximation (compare also this approach with the ones contained in [26-27]) well known and often used VdW and Dieterici EOS has been adopted in asymptotic form which has been correlated with the method based on symmetries of the first-order derivatives pressure and chemical potential with respect to temperature [25] (for more details see Sec.2). In this method, the authors couple some type of Maxwell's construction with first–order nonlinear differential equations theory. Taking into account the fact that differential- and integral equation theories are quite successful in predicting the thermodynamic and structural properties of a variety of simple and more complex systems over a wide domain in temperature-density space [25, 27-28], they meet with variable success in the critical region due to an unsatisfactory treatment of long wavelength fluctuations which are of particular relevance in this region [29-34]. We expand described in [25] and modified in [26] method of some simple differential theory for two-dimensional systems with singularities.

## 2. Main assumptions and conditions of the presented method

Derived in [25] very simple way to attain differential form of the vapor-liquid equilibrium curve as well as thermodynamic properties of the VLE system, is based on the assumption that for any temperature $T$ below the critical $T_c$, the curves $\rho_L = \rho_L(T)$ for liquid and $\rho_V = \rho_V(T)$ for vapor, satisfy the following system of equations:

$$P(\rho_V(T), T) = P(\rho_L(T), T), \quad \mu(\rho_V(T), T) = \mu(\rho_L(T), T) \tag{4}$$

Both the chemical potential, $\mu$ and the pressure, $P$ are differentiable functions of density, $\rho$ and temperature, $T$ hence differentiating (4) with respect to $T$ leads to

$$\frac{\partial P}{\partial T}(\rho_V(T), T) + \frac{\partial P}{\partial \rho}(\rho_V(T), T)\rho_V'(T) = \frac{\partial P}{\partial T}(\rho_L(T), T) + \frac{\partial P}{\partial \rho}(\rho_L(T), T)\rho_L'(T)$$
$$\frac{\partial \mu}{\partial T}(\rho_V(T), T) + \frac{\partial \mu}{\partial \rho}(\rho_V(T), T)\rho_V'(T) = \frac{\partial \mu}{\partial T}(\rho_L(T), T) + \frac{\partial \mu}{\partial \rho}(\rho_L(T), T)\rho_L'(T) \tag{4a}$$

where $\rho_L'$ and $\rho_V'$ denotes the first-order derivatives of $\rho_L(T)$ and $\rho_V(T)$ with respect to $T$.
Equation (4a) defines a system of two first- order nonlinear differential equations given by differential form of the curves

$$\rho_L'(T) = f(\rho_L(T), \rho_V(T), T) \text{ and } \rho_V'(T) = f(\rho_V(T), \rho_L(T), T) \tag{4b}$$

where function $f$ is the form

$$f(\rho_V, \rho_L, T) = \frac{\left(\frac{\partial \mu}{\partial T}(\rho_V, T) - \frac{\partial \mu}{\partial T}(\rho_L, T)\right)\rho_V\rho_L + \left(\frac{\partial P}{\partial T}(\rho_L, T) - \frac{\partial P}{\partial T}(\rho_V, T)\right)\rho_V}{\frac{\partial P}{\partial \rho}(\rho_V, T)(\rho_V - \rho_L)} \tag{4c}$$

Taking into account the fact that the right side of the definition (4c) is regular for $T < T_c$ and $\rho_V(T) < \rho_L(T)$, we can find a unique local solution in the form of curves $\rho_L$ and $\rho_V$ for given initial conditions $\rho_V^0 = \rho_V(T_0)$, $\rho_L^0 = \rho_L(T_0)$, $\rho_V^0 < \rho_L^0$ and temperature $T_0 < T_c$. Additionally, it is assumed that $\rho_V(T_c) = \rho_L(T_c) = \rho_c$ (where $(\rho_c, T_c)$ determines the so-called critical point and $\rho_c$ and $T_c$ denote the density and critical temperature, respectively).
Usually it is almost impossible to appoint a strict, smooth solution of differential equation (4b) with (4c), for given $P$ and $\mu$, and numerical method must be used [21, 35-38]. We modified method proposed in [25] correlating it with the quantitative description of the phenomena associated with the phase transitions. As it turns out this method can be successfully used to calculate the critical exponent value in the asymptotic vicinity of the VLE system.
In our approach we do not assume analyticity of $f$, $P$, $\mu$ functions, we require them only to be smooth enough differentiable up to some order.

## 3. Results and discussion

A new mathematical way, derived in this paper, to determine the value of the critical exponent for the VLE system is based on some type of classical assumptions – *mean field theory* as well as asymptotic form very simple and commonly used EOS's for real gases. In like manner (as in the case of classical assumptions) our method contain both the contribution of any external field and the effect on a given particle of the remaining ones. Moreover, the disregard of fluctuations (inherent to mean-field theories) is also taken into account.

In Section 3.1 we consider Van der Waals EOS, while in Section 3.2 Dieterici model is considered. Finally in Section 3.3 we cover the case of alternative mathematical approach with different approximation scheme.

### 3.1 Critical exponent theory versus symmetrical properties of the derivatives of the densities in the vapor and liquid phases –an analytical approach (case of the Van der Waals EOS)

From mathematically straightforward method used previously to determine the VLE curves in parabolic and cubic shape, in this paper we modified this manner so that it was possible to determining the value of the critical exponent of the VLE system. In fact we determine the reduced densities of the liquid and vapor in the neighborhood of critical points, cf. subsequently formulas (25) and (26).
In the first approach we used the most popular VdW EOS [26] for real gases in the form

$$P(V,T) = \frac{RT}{V-b} - \frac{a}{V^2}$$

where $P$ – pressure, $V$ – volume, $R$ – universal gas constant, $a$ - represents attraction arising from dispersion forces, $b$ - accounts for the volume occupied by the molecules.
Substituting the relation $\rho = m/V = 1/V$ we obtain, abusing the notation, that

$$P(\rho,T) = \frac{\rho RT}{1-\rho b} - a\rho^2 \tag{5}$$

We consider the following VdW EOS in the reduce variables (see also the note at the equation (1) for the case of the critical density as well as Eqs. (5) and (7) in work [26])

$$p(\tilde{\rho},t) = \frac{8(t+1)(\tilde{\rho}+1)}{(2-\tilde{\rho})} - 3(\tilde{\rho}+1)^2 - 1 \tag{6}$$

Because we used in the calculation the reduced units $p$, $\rho$, $\mu$ which are related to the original ones and which can be defined as $\mu_c(\mu + 1)$, $\rho_c(\rho + 1)$, $p_c(p + 1)$ – scaled by the values of the critical parameters, where $\mu_c$, $\rho_c$, $p_c$ are critical values of the chemical potential, density and pressure, respectively (see also note in Sec.2), it can be seen that the below thermodynamic relation between $P$ and $\mu$, is slightly differ (by using the denominator ($\rho$ +1) instead of $\rho$) from the ones adopted for not reduced variables (comments: the relation for the constants $p_c = \rho_c \mu_c$ also holds, according to the corresponding states theorem, see note in Introduction and [10]),

$$\frac{\partial \tilde{\mu}}{\partial \tilde{\rho}} = \frac{1}{\tilde{\rho}+1} \frac{\partial p}{\partial \tilde{\rho}} \tag{7}$$

The approximate form of the chemical potential, designating chemical thermodynamic equilibrium in the system, we get almost directly from the above thermodynamic relation

$$\tilde{\mu}(\tilde{\rho},t) = 8(t+1)\left(\frac{1}{3}ln\left(\frac{\tilde{\rho}+1}{2-\tilde{\rho}}\right) - \frac{1}{2-\tilde{\rho}}\right) - 6\tilde{\rho} + \tilde{C}(t) \tag{7a}$$

Above formulas one can derive from the necessary expressions for first derivatives of the pressure and chemical potential with respect to temperature and pressure with respect to density (see also right side of equation (4c))

$$\frac{\partial p}{\partial t} = 8\frac{\tilde{\rho}+1}{2-\tilde{\rho}}; \qquad \frac{\partial p}{\partial \tilde{\rho}} = 24\frac{(t+1)^2}{2-\tilde{\rho}} - 6(\tilde{\rho}+1); \qquad \frac{\partial \tilde{\mu}}{\partial t} = \frac{8}{3}ln\left|\frac{\tilde{\rho}+1}{2-\tilde{\rho}}\right| - \frac{8}{2-\tilde{\rho}}$$

And next using the approximations $ln(z + 1) \sim z$ valid for $z$ small enough we can approximate with $r = \tilde{\rho}_L - \tilde{\rho}_V$ the expression

$$\frac{8}{3}ln\left(1 + \frac{3r}{(\tilde{\rho}_L+1)(\tilde{\rho}_V-2)}\right) \sim \frac{8}{3}\frac{3r}{(\tilde{\rho}_L+1)(\tilde{\rho}_V-2)} \sim 4r$$

From (4c) and (5) by using (6) and (7a), after some mathematical modification (compare also with [27]), we can derived as follows

$$\tilde{\rho}'_L = \frac{-8\,(\tilde{\rho}_L+1)(1+\tilde{\rho}_V)(2-\tilde{\rho}_L)}{12(t+1)(2-\tilde{\rho}_V)-3(\tilde{\rho}_L+1)(2-\tilde{\rho}_L)^2(2-\tilde{\rho}_V)}, \quad \tilde{\rho}'_V = \frac{-8\,(\tilde{\rho}_V+1)(1+\tilde{\rho}_L)(2-\tilde{\rho}_V)}{12(t+1)(2-\tilde{\rho}_L)-3(\tilde{\rho}_V+1)(2-\tilde{\rho}_V)^2(2-\tilde{\rho}_L)} \qquad (8)$$

Now we want to show that from nonlinear differential equations (8) we can conclude/ obtain the value of the critical exponent (critical index), which in the near – critical point depends on the temperature as $|T - T_c|^{\pm\varepsilon} \cong |T/T_c - 1|^{\pm\varepsilon} \cong B_0 |t|^{\pm\varepsilon}$, where $\varepsilon > 0$ is called an index or a critical exponent, $B_0$ is called critical amplitude, in this case (see also abbreviations in Tab. 1).

At the beginning system of nonlinear differential equations (8) we can rewrite in the following form

$$\frac{12(t+1)\tilde{\rho}'_L}{(\tilde{\rho}_L+1)(2-\tilde{\rho}_L)} = \frac{8\,(\tilde{\rho}_V+1)}{(\tilde{\rho}_V-2)} + 6\tilde{\rho}'_L - 3\tilde{\rho}_L\,\tilde{\rho}'_L \qquad (9)$$

and similarly in symmetric way

$$\frac{12(t+1)\tilde{\rho}'_V}{(\tilde{\rho}_V+1)(2-\tilde{\rho}_V)} = \frac{8\,(\tilde{\rho}_L+1)}{(\tilde{\rho}_L-2)} + 6\tilde{\rho}'_V - 3\tilde{\rho}_V\,\tilde{\rho}'_V \qquad (10)$$

After introducing auxiliary functions $H'(\tilde{\rho})(\tilde{\rho}+1)(2-\tilde{\rho}) = 1$ and $(\tilde{\rho}-2)G(\tilde{\rho}) = 8(\tilde{\rho}+1)$ hence $3H(\tilde{\rho}) = \ln\left(\frac{1}{8}|G(\tilde{\rho})|\right)$ we rephrase (9) as follows

$$12(t+1)\tilde{\rho}'_L \dot{H}(\tilde{\rho}_L) = G(\tilde{\rho}_V) + 6\tilde{\rho}'_L - 3\tilde{\rho}_L\,\tilde{\rho}'_L \qquad (11)$$

And similarly equation (10) can be reformulated as

$$12(t+1)\tilde{\rho}'_V \dot{H}(\tilde{\rho}_V) = G(\tilde{\rho}_L) + 6\tilde{\rho}'_V - 3\tilde{\rho}_V\,\tilde{\rho}'_V \qquad (12)$$

Taking the sum and the difference of the equations (11) and (12) one obtains the equations for the sum and the difference between two phases (sum and difference between two phases is denoted as $s = \tilde{\rho}_V + \tilde{\rho}_L$ and $r = \tilde{\rho}_L - \tilde{\rho}_V$, moreover we must mention in this place that the *s* parameter, in contrast to the *r* parameter, which is called the order parameter of the VLE system, do not have any physical meaning and has been introduced for the sake of brevity in notation), which are the following

$$12(t+1)\bigl(H(\tilde{\rho}_L) + H(\tilde{\rho}_V)\bigr) - 6s' + \tfrac{3}{4}(s^2 + r^2)' = G(\tilde{\rho}_L) + G(\tilde{\rho}_V) \qquad (13)$$

and

$$12(t+1)\bigl(H(\tilde{\rho}_L) - H(\tilde{\rho}_V)\bigr) - 6r' + \tfrac{3}{2}(rs)' = G(\tilde{\rho}_L) - G(\tilde{\rho}_V) \qquad (14)$$

To reduce singularity of the order one in *r* (as is continue in Sec. 3.3) from (14) insert $(t + 1)$ into (9)-(10). From equation (13) approximating

$$H(\tilde{\rho}_L) + H(\tilde{\rho}_V) = \tfrac{1}{3}\ln\tfrac{1}{4} + \tfrac{1}{2}s \qquad (15)$$

up to the order $s^2$ for *s* small enough and

$$G(\tilde{\rho}_L) + G(\tilde{\rho}_V) = -8 - 2s + s^2 - r^2 \qquad (16)$$

we obtain from (13), up to *ss'* order, that

$$6ts' + \tfrac{3}{4}(s^2 + r^2)' = -8 - 2s + s^2 - r^2 \qquad (17)$$

Hence for $s = A\,(-t)^{\tau}$, $r = B\,(-t)^{\beta}$, $\beta < \min\{\tau, (\tau+1)/2\}$ (where *B* is correlated with the critical amplitude value of the VLE system, see also Introduction and References therein) we considering the leading terms of the zero order in (-*t*) for $2\beta - 1 = 0$, ¾ $(r^2)'$ and (-8) (we must comment in this place that the asymptotical form of the *s* parameter does not correspond to any critical exponent value of the VdW model for the vapor – liquid phase transition and any critical amplitude of these system. This has been introduced only for the sake of brevity in notation as well as to simplify the mathematical calculations) and finally we get

$$3rr' = -16 \tag{18}$$

And after integration

$$r = \sqrt{-t\frac{32}{3}} \tag{19}$$

Therefore, the critical exponent $\beta$ is equal ½ and $\tau$ should be greater than ½ while $B = 4\sqrt{6}/3$.

Without postulated approximation of $r$ and $s$, if we assume that $s' > 0$ and $s \in [0, 2]$ then from (17) we get

$$6rr' \leq -32 \tag{20}$$

hence, after integration, that for any $t \leq 0$, thus justifying one sided estimate in (19)

$$r^2(t) \leq -\frac{32}{3}t \tag{21}$$

Approximating now $G(\tilde{\rho}_L) - G(\tilde{\rho}_V) = -6r$ and $H'(\tilde{\rho}_L) - H'(\tilde{\rho}_V) = \frac{r}{2}$ forgetting in the last difference all the terms of order less than $r$ we obtain from (14), analogously to the relation derived from the sums, *i.e.* (17), the following relation

$$12tr' + 3r's + 3rs' = -6r \tag{22}$$

Then if we use once again the approximating solutions $r = B(-t)^\beta$ and $s = A(-t)^\tau$, then the assumptions that $rs'$ is negligible in comparison with other terms leads to $\beta < 1$ and yields $\tau = 1$ which contradicts with our previous derivation $\tau = ½$. But if we suppose that $\beta = 1$ then the relation (22) gives

$$6 = 6\tau + \frac{3}{2}B(\tau + \beta) \tag{23}$$

hence $B = 4/3$ follows if we use the values of $\tau = 1/2$ and $\beta = 1$. Summarizing we thus obtained

$$s = -\frac{4}{3}t \tag{24}$$

Therefore using the formulas for $r$ and $s$ we can derive the approximate formulas for small values of $t \leq 0$, for $\tilde{\rho}_L$ and $\tilde{\rho}_V$ which reads

$$\tilde{\rho}_L(t) = \frac{s(t) + r(t)}{2} = \frac{1}{2}(-\frac{4}{3}t + \sqrt{-t\frac{32}{3}} + o(\sqrt{-t})) \tag{25}$$

And respectively

$$\tilde{\rho}_V(t) = \frac{s(t) - r(t)}{2} = \frac{1}{2}(-\frac{4}{3}t - \sqrt{-t\frac{32}{3}} + o(\sqrt{-t})) \tag{26}$$

Note that in the $o(\sqrt{-t})$ can be the terms *e.g.* of order one making the expression for $\tilde{\rho}_V$ positive in the neighborhood of zero and changing thus of coefficient *4/3* at the first power of *t* in both expression for $\tilde{\rho}_L$ and $\tilde{\rho}_V$ though not for *s* as in (24).

## 3.2 Critical exponent theory versus symmetrical properties of the derivatives of the densities in the vapor and liquid phases –an analytical approach (case of the Dieterici EOS)

To the proved that the value of the critical exponent for the VLE system determined from presented above straightforward mathematical method based on simple differential equation theories and some kind of the Maxwell's construction, is correct and valid, we consider another EOS, i.e. *Dieterici* EOS, which is the modification of the VdW EOS. To be more specific we formulate the formulas of difference the system of EOS's for Dieterici case, cf. (30a).
Analogously as in the Sec. 3.1 the *Dieterici* EOS in the reduced variables *p, ρ, μ* has the form

$$p(\tilde{\rho}, t) = \frac{(t+1)(\tilde{\rho}+1)}{(1-\tilde{\rho})} exp\left[2\left(\frac{t-\tilde{\rho}}{t+1}\right)\right] - 1$$

Hence after using the approximation $exp(z) \sim 1 + z + \frac{1}{2}z^2$ we obtain

$$p(\tilde{\rho}, t) = \frac{3t(\tilde{\rho}+1)}{(1-\tilde{\rho})} + \frac{(\tilde{\rho}+1)(1-2\tilde{\rho})}{(1-\tilde{\rho})} + \frac{2(\tilde{\rho}+1)(t-\tilde{\rho})^2}{(1-\tilde{\rho})(t+1)} - 1 \qquad (27)$$

so that the crucial properties of the pressure function *p* survive the approximation

$$\frac{\partial p}{\partial \tilde{\rho}}_{(0,0)} = \frac{\partial^2 p}{\partial \tilde{\rho}^2}_{(0,0)} = 0$$

The approximate form of the chemical potential, designating chemical thermodynamic equilibrium in the system, we get almost directly from the mentioned in Sec. 3.1 thermodynamic relation (7)

$$\tilde{\mu}(\tilde{\rho}, t) = \frac{3}{2}(t+1)ln(1+\tilde{\rho}) - \frac{1}{2}(3t-1)ln(1-\tilde{\rho}) + \frac{1-3t}{(1-\tilde{\rho})} + \tilde{C}(t) \qquad (28)$$

As it was pointed out in Sec. 3.1 the above approximations one can derive almost directly from the expressions for first derivatives of the pressure and chemical potential with respect to temperature, pressure with respect to density and by using the thermodynamic relation - in reduced variables, between *P* and *μ*, is (see also the right side of equation (4c) and Sec. 3.1)

$$\frac{\partial p}{\partial t} = \frac{(\tilde{\rho}+1)}{1-\tilde{\rho}}\left(3 + 2\frac{(t-\tilde{\rho})(t+2+\tilde{\rho})}{(t+1)^2}\right); \qquad \frac{\partial p}{\partial \tilde{\rho}} = \frac{6t}{(1-\tilde{\rho})^2} + \frac{2\tilde{\rho}(\tilde{\rho}-2)}{(1-\tilde{\rho})^2} + \frac{4(t-\tilde{\rho})(t+\tilde{\rho}^2-\tilde{\rho}-1)}{(t+1)(1-\tilde{\rho})^2}$$

Then, up to some term of the *t* variable, with $(1 - \tilde{\rho})^2 H'(\tilde{\rho}) = 4(\tilde{\rho}^2 - \tilde{\rho} - 1) - 4\tilde{\rho}/(\tilde{\rho}+1)$

$$\frac{\partial \tilde{\mu}}{\partial t} = \left(\frac{3}{2} + \frac{t}{2(t+1)^2}\right)ln\left|\frac{\tilde{\rho}+1}{1-\tilde{\rho}}\right| + \frac{3-2t/(t+1)^2}{\tilde{\rho}-1} + \frac{H(\tilde{\rho})}{(t+1)^2}$$

The above partial derivatives we can using to get the approximated, with the exponent function replaced by the first two terms, *i.e.* quadratic approximation, Dieterici model. Note that $H(\tilde{\rho}) \sim \tilde{\rho}^4 - 6\tilde{\rho}^2 - 4\tilde{\rho}$ in a power like approximation.
And next using the first term of Taylor's expansion for the logarithm function $ln(z+1) \sim z$ with $z = 2\tilde{\rho}/(1-\tilde{\rho})$ we can approximate the expression

$$ln\left(\frac{(\tilde{\rho}_L+1)(1-\tilde{\rho}_V)}{(1-\tilde{\rho}_L)(1+\tilde{\rho}_V)}\right) \sim 2r = 2(\tilde{\rho}_L - \tilde{\rho}_V)$$

From (4c) and (5) by using (27) and (28), we can derive (for some $\tilde{B}$) as follows

$$\tilde{\rho}'_L = \frac{\tilde{B}+8t}{6t-4\tilde{\rho}_L}, \quad \tilde{\rho}'_V = \frac{\tilde{B}+8t}{6t-4\tilde{\rho}_V} \qquad (29)$$

Just it was done in Section 3.1 we use a procedure based on the introduction of the definition of auxiliary functions $\dot{H}(\tilde{\rho})$ as well as the *s* and *r* parameters, for the sake of brevity in notation, we obtain the following relations

$$6(ts)' - (s^2 + r^2)' = 16t + 2\tilde{B} \quad \text{and} \quad 3(tr)' = 2r's' \tag{30}$$

It is worth to noting that in above equation for $\beta < \min\{\tau, (\tau+1)/2\}$ the leading terms are $(-r^2)'$ and $(2\tilde{B})$ of the order $2\beta - 1 = 0$, while the other terms are of the higher order $(\tau)$ and $(2\tau - 1)$. Hence if we putting for $s = A(-t)^\tau$, $r = B(-t)^\beta$, and $\beta < \min\{\tau, (\tau+1)/2\}$ we get directly that

$$r^2 = -2\tilde{B}t \quad \text{and} \quad r = \sqrt{-2\tilde{B}t} \tag{30a}$$

So that in this case is also $\beta = \frac{1}{2}$ and value of the critical amplitude depends on value of constant $\tilde{B}$, i.e. $B = \sqrt{2\tilde{B}}$.

We proved that the value of the critical exponent for the VLE system determined by using presented above straightforward mathematical method based on simple differential equation theories and some kind of the Maxwell's construction, is correct and valid, for another EOS, i.e. *Dieterici* EOS (as in the case of the VdW EOS). This fact is agreement with the classical assumptions, because we taking into account both, the contribution of any external field and the effect on a given particle of the remaining ones. We also must mention in this place that the neglect of fluctuations (inherent to mean-field theories) is also included.

**3.3 Critical exponent theory versus symmetrical properties of the derivatives of the densities in the vapor and liquid phases – an alternative academic approach**

Firstly it must be underlined in this place, that the results obtained in this Section are irrelevant for the previous cases considered in Sections 3.1 and 3.2 since the formula (31) is the generalized academic approach.
So we reflect an alternative differential expression of the VLE curves based on simple, asymptotic form of the academic EOS's (covering approximate *VdW* and Dieterici cases, that can be considered as simple and predictive, because they contain only two fixed parameters, calculated from the critical properties), raising some interesting questions. It is well known that most of the existing EOS's are based on different combinations and choices for the repulsive and attractive terms and developed following the *VdW* scheme. The primary reason for the development of a new EOS's is the significant improvement in results for the VLE properties. In most cases, the proposed EOS's are cubic in volume, so that the VdW idea is preserved.
As shown in Sec.3.1 and 3.2, simple mathematical structure differential expression of the coexistence densities curves allows one to calculate not only some thermodynamic properties analytically [26-27] but also both proved and derive, the value of the critical exponent in the case when the vapor – liquid critical point is achieved asymptotically along the coexistence curve.
We consider a nonautonomous, singular system of differential equations. Our approach is to reduce the singularity of cancellation, i.e. both the nominators and denominators appearing on the right hide side of the system share the same asymptotic close to the critical value of the temperature thus allowing us to skip this difficulty in dealing with such singular systems. Nevertheless some singularity might still be present in the system which is the case when the value of the parameter *C* is equal to zero like in the paper [39] thus allowing for singular in the derivative solutions to our system.
To seeking deeper explanation of the behavior differential expression of the coexistence densities curves we take into consideration in this Section the general quadric approximation model, i.e. model in which the analytical expression of these VLE curves are defined for any $t \in [t_0, t^*]$ and constants *C*, *D* as

$$\tilde{\rho}_L' = \frac{\tilde{\rho}_L - \tilde{\rho}_V}{2t + C\tilde{\rho}_V^2 + D\tilde{\rho}_V}, \quad \tilde{\rho}_V' = \frac{\tilde{\rho}_V - \tilde{\rho}_L}{2t + C\tilde{\rho}_L^2 + D\tilde{\rho}_L} \tag{31}$$

The above system can be considered as the generalization of the mathematical model of the phase transition presented in [39] with *C* = 6 and *D* = 0, of the approximation of the *VdW* model with *C* = 3/2 and *D* = –4 [26-27] and finally the approximation of the *Dieterici* model for *C* = 2/3 and *D* = 4/3, adopted formulas (14) an (31) from [26].
The model (31) and the derivation below are not related to the results from Sections 3.1and 3.2.

*DEFINITION 1* By a solution of the system (31) we mean a pair of the density respectively vapor and liquid functions of the temperature variable $t \in [t_0, t^*]$ denoted by ($\rho_V(t)$, $\rho_L(t)$), such that $\rho_V, \rho_L \in C^1([t_0, t^*]) \cap C([t_0, t^*])$, $\rho_V(t^*) = \rho_L(t^*)$ for some $t^* > t_0$, while $0 \leq \rho_V(t) < \rho_L(t)$ for any $t \in [t_0, t^*]$ and finally $|\rho_V'(t^*)| = |\rho_L'(t^*)| = \infty$.

*LEMMA 1* If $(\rho_V, \rho_L)$ are solutions of the system in terms of Definition 1 then for any $t \in [t_0, t^*]$

$$h(\rho_V(t)) - h(\rho_L(t)) + 2t(\rho_V(t) - \rho_L(t)) = 0 \tag{32}$$

where $h(\rho) = \frac{1}{3}C\rho^3 + \frac{1}{2}D\rho^2$.

**Proof.** Take the difference of the equations from the system (31) and integrate using $\rho_V(t^*) = \rho_L(t^*)$ the following thus obtained relation for any $t \in [t_0, t^*]$

$$\rho_V'(t)(2t + H\rho_V(t)) - \rho_L'(t)(2t + H(\rho_L(t))) = 2(\rho_L(t) - \rho_V(t))$$

where $H(\rho) = C\rho^2 + D\rho$ ∎

*LEMMA 2* The *Lemma 1* can be rephrased for any $t \in [t_0, t^*]$ as

$$4t + D(\rho_L(t) - \rho_V(t)) + \frac{2}{3}C(\rho_V^2(t) + \rho_V(t)\rho_L(t) + \rho_L^2(t)) = 0 \tag{33}$$

or equivalently, if we introduce new variables (see also Sec. 3.1 and 3.2) $s = \tilde{\rho}_V + \tilde{\rho}_L$ and $r = \tilde{\rho}_L - \tilde{\rho}_V$, as

$$Cr^2 = 24t - 6Ds - 3Cs^2. \tag{34}$$

The above conservation law (33) from *Lemma 2* allows us to reformulate the system (31) in a slightly, of order in $r$, desingularised and autonomized form

$$\tilde{\rho}_L' = \frac{6}{3D + 2C(2\tilde{\rho}_L + \tilde{\rho}_V)}, \quad \tilde{\rho}_V' = \frac{-6}{3D + 2C(2\tilde{\rho}_V + \tilde{\rho}_L)} \tag{35}$$

*LEMMA 3* If $\rho_V(t^*) = \rho_L(t^*) = \rho^*$ for solution $(\rho_V, \rho_L)$ of (31) defined by the *Definition 1* then $\rho^* = -D/2C$ and $t^* = -D^2/8C$. Moreover, if $t^* = 0$ then $\rho^* = 0$ and $D = 0$.

**Proof.** Indeed, it is sufficient to calculate one of the denominators of (35) substituting $\rho_V(t^*) = \rho_L(t^*) = \rho^*$ and thus getting the necessary condition for the phase transition

$$\rho^* = \frac{-D}{2C} \tag{36}$$

Similarly the denominators of (31) at $t^*$, by *Definition 1*, necessarily equals zero

$$2t^* + C(\rho^*)^2 + D\rho^* = 0.$$

Thus from the above or (33) – (34) combined with (36) one gets

$$t^* = \frac{D^2}{8C} \quad ∎ \tag{37}$$

Therefore, if we consider $t$ to be the reduced temperature variable we thus have obtained in the only possibility of $D = t^* = 0$. Otherwise the general system of the form (31) allows for singular solutions, according to the definition (1), only at the point $t^* = D^2/8C$ with the corresponding values of the $C$ and the $D$ parameters. It should be underlined that the most relevant situation is for $D = 0$ *i.e.* the system introduced in [39] as the approximation of some important physical system allowing for the phase transition.

## 4. Concluding remarks

This paper presents a very simple, from mathematical point of view, method that allows the rigorous justification for the critical exponent value of the density curves for both phases and the same the description of the critical region for the determination of VLE system. The value of the critical exponent, in the case when the vapor – liquid critical point is achieved asymptotically along the coexistence curve, is in effect correlating both, some type of Maxwell's construction with first–order nonlinear differential equations theory and chosen EOS's for real gases. This approach provides the specific value of the critical exponent equal to that based on classical assumptions (*i.e.* mean-field theory). Moreover based on postulated method one can estimate value of the critical amplitude in general case (*i.e.* in the case when we do not take into consideration any pure compounds or mixtures as in the refractive index gradient method [10] and References therein). What's more 'non universal', *i.e.* not depended on the system analyzed and the substances tested, value of the critical amplitude $B_0$ (for the VLE system) can't be obtained from the RG theory only for the Ising system with large distinct nearest neighbor interaction operator (*i.e.* for large value of *s*). Taking into account the nonanalytic nature of the critical region the large difference between the value of the critical amplitude determined in this work and those based on the results obtained via experimental efforts, is a consequence of the fact, that we excluded in our method the effect of molecular shape, other complicating factors, the influence of gravitational effects or the influence of dipole interactions. Hence the value of the critical amplitude determined from VdW EOS's (see also Sec. 3.1 where the $B \sim 3.266$) is nearly two times greater than this for selected substances of some diatomic and polyatomic structures (compare also this result with those contained in Tab. 3). We expect such large discrepancy between the results obtained by adopting the VdW and Dieterici EOS in calculation than these contained in work [10]. It's obvious that the VdW EOS and its derivation (as for example Dieterici EOS) much better reflects the nature of the theoretical calculations than practical applications.

In Sec. 3.3 we have reconsidered differential expression of the VLE curves (as a case of nonautonomous, singular system of differential equations) based on simple, asymptotic form of the academic EOS's (including *VdW* and Dieterici cases), raising some interesting questions. In this Section we proved that for only one possibility with $D = t^* = 0$ the system introduced in [39] as the approximation of some important physical system allows for the phase transition, take place the most relevant situation.

We hope that this model can be useful for additional studies of the VLE system and we intend to publish additional correlations for other simply fluids, using method similar to those used in this work.

## Tables

**Table 1** Power laws, corresponding critical exponents and critical amplitudes as well as the ways of achieving the critical point for vapor - liquid system.

| Power laws | Critical exponent | Trajectory |
|---|---|---|
| $\tilde{C}_v^\pm = A_0^\pm |t|^{-\alpha}$ | $\alpha$ | critical isochore, $\tilde{\rho} = 0$ |
| $\tilde{\rho} = B_0 |t|^\beta$ | $\beta$ | coexistence curve, $\tilde{\rho} = \tilde{\rho}_{coex}$ |
| $\tilde{\chi}_T = \Gamma_0^+ |t|^{-\gamma}$ | $\gamma$ | critical isochore, $\tilde{\rho} = 0, t \geq 0$ |
| $\tilde{\chi}_T = \Gamma_0^- |t|^{-\gamma}$ | $\gamma$ | coexistence curve, $\tilde{\rho} = \tilde{\rho}_{coex}$ |
| $\tilde{\mu} = D_0 \tilde{\rho}|\tilde{\rho}|^{\delta-1}$ | $\delta$ | critical isotherm, $t = 0$ |
| $p = D_0 \tilde{\rho}|\tilde{\rho}|^{\delta-1}$ | $\delta$ | critical isotherm, $t = 0$ |
| $\zeta = \zeta_0^+ |t|^{-\nu}$ | $\nu$ | critical isochore, $\tilde{\rho} = 0, t \geq 0$ |
| $\zeta = \zeta_0^- |t|^{-\nu}$ | $\nu$ | coexistence curve, $\tilde{\rho} = \tilde{\rho}_{coex}$ |

Abbreviations:
$t = \frac{T}{T_c} - 1$ ; $\tilde{\rho} = \frac{\rho}{\rho_c} - 1$; $\tilde{\mu} = \frac{\mu}{\mu_c} - 1$; $p = \frac{P}{P_c} - 1$; $A_0^\pm; B_0; D_0; \Gamma_0^\pm; \zeta_0^\pm$ - critical amplitudes

**Table 2** Values of the critical exponent for the VLE system in the case of the 3D Ising model and mean field approaches.

| Critical exponent | Based on [40] | Based on [41] | Mean field approaches |
|---|---|---|---|
| $\alpha$ | 0.110±0.003 | 0.110 (1) | 0 |
| $\beta$ | 0.326±0.002 | 0.3265 (3) | 0.5 |
| $\gamma$ | 1.239±0.002 | 1.2372 (5) | 1 |
| $\nu$ | 0.630±0.001 | 0.6301 (4) | 0.5 |
| $\delta$ | 4.80±0.02 | 4.789 (2) | 3 |

**Table 3** Selected values of the critical amplitude for the VLE system (power law for the order parameter $\rho_L - \rho_V$) in the case of the simple Lennard-Jones fluids.

| Molecule/ fluid | Critical amplitude $B_0$ Based on [10] |
|---|---|
| $^4$He | 1.100±0.006 |
| Xe | 1.426±0.009 |
| $C_2H_6$ | 1.564±0.008 |
| $CO_2$ | 1.636±0.009 |
| $N_2O$ | 1.598±0.009 |